\documentstyle[aps,prl,graphics]{revtex}

\begin{document}
\twocolumn[\hsize\textwidth\columnwidth
\hsize\csname@twocolumnfalse\endcsname
\draft
\title {Universal Heat Conduction in YBa$_2$Cu$_3$O$_{6.9}$}
\author {Louis Taillefer, Benoit Lussier$^*$ and Robert Gagnon}

\address {Department of Physics,
McGill University,
Montr\'eal, Qu\'ebec, Canada H3A 2T8}

\author {Kamran Behnia and Herv\'e Aubin}

\address {Laboratoire de Physique des Solides (CNRS),
Universit\'e Paris-Sud, 91405 Orsay, France}

\maketitle

\date{10 March 1997}

\begin {abstract}

The thermal conductivity of
YBa$_2$Cu$_3$O$_{6.9}$ was measured at low temperatures
in untwinned single crystals with concentrations of Zn impurities
from 0 to 3\% of Cu.
A linear term $\kappa_{0}/T=0.19$~mW~K$^{-2}$~cm$^{-1}$
is clearly resolved as $T \rightarrow 0$,
and found to be virtually
independent of Zn concentration. The existence of this residual normal fluid
strongly validates the basic theory of transport in unconventional
superconductors.
Moreover, the observed universal behavior is in
quantitative agreement
with calculations for a gap function of $d$-wave symmetry.

\end{abstract}


\vskip1pc]

\narrowtext

The theory of quasiparticle transport in unconventional superconductors,
developed over the last decade, has remained largely untested.
A novel feature that arises when the superconducting gap function
has nodes for certain
crystal directions
is the existence of quasiparticles at $T=0$.
This residual normal fluid
is a consequence of impurity scattering, even for low concentrations of
non-magnetic
impurities
(see \cite{HVW,Varma}, and references therein).
Its presence, which should dominate the conduction of
heat and charge at $T<<T_c$, has yet to be firmly established, and its
properties have never
been investigated.
For certain pairing states, with appropriate gap topology and symmetry,
an appealing phenomenon is predicted to occur:
quasiparticle transport should be independent of scattering rate as
$T\rightarrow0$.
This universal limit, first pointed out by Lee \cite{Lee}
for the case of a $d$-wave gap in two dimensions, is the result of a
compensation
between the growth in
normal fluid density with increasing impurity concentration and
the concomitant reduction in mean free path.

In this Letter, we report the first observation of universal transport in a
superconductor. Our study of heat conduction in the high-$T_c$ cuprate
YBa$_2$Cu$_3$O$_{6.9}$ provides a solid validation of the
basic theory of transport in unconventional superconductors
and insight into the nature of impurity scattering in the cuprates.
It also supports strongly an identification of the gap function as having
$d$-wave symmetry.

The thermal conductivity $\kappa(T)$ was measured between
0.05 and 1~K, for a current along the $a$ axis of
five single crystals: four untwinned crystals of
YBa$_2$(Cu$_{1-x}$Zn$_x$)$_3$O$_{6.9}$
and one
crystal of YBa$_2$Cu$_3$O$_{6.0}$. The latter was obtained by full
deoxygenation via
annealing at 800~$^{\circ}$C in helium gas for 64~h; it is insulating, with
$\rho_a$(100~K)~=~42.7~$\Omega$~m.
$x$ is the nominal concentration of Zn, achieved by mixing in
ZnO powder at the start of
the growth process in the atomic ratio Zn:Cu::$1.5x:1-x$, for $x=0$, 0.006,
0.02 and 0.03.
The experimental technique and the sample preparation are described
elsewhere \cite{Chains,Gagnon}.
The
resistive
$T_c$ is given in Table~I.
The uncertainty on
the geometric factor is
at most $\pm$~10\%, 10\%, 20\%, 5\% and
10\%
for the $x=0$ (``pure''), 0.6\%, 2\%, 3\% and deoxygenated (``deox'')
samples, respectively.

The $a$-axis resistivity is linear in temperature
above 130~K
\cite{Gagnon}, and a fit to $A+BT$ yields the values in Table~I.
Zn substitution has two effects:
it reduces $T_c$ and it increases $A$.
At low concentration, both effects are linear, and
$dT_c/dA=-~0.5$~K~/~$\mu\Omega$~cm, in agreement with data on twinned crystals
(e.g.
\cite{Fukuzumi}).
Concentrations of Zn from 0 to 3\% correspond
to a large range of scattering rates, but
to a modest level of pair-breaking:
adding
3\% Zn
suppresses
$T_c$
by only
20\%.
Given that the inelastic scattering term $B$ is independent of $x$,
the impurity scattering rate $\Gamma=1/(2\tau_0)$
may be estimated via the residual resistivity
$\rho_0=m^*/ne^2\tau_0$:

\begin{equation}
\Gamma_{\rho}(x)=(\omega_p^2/8\pi)[~\rho_0(x=0)+A(x)-A(0)~]
\end{equation}

\begin{table}[hb]
\caption[]{Sample characteristics for the four untwinned a-axis
crystals of YBa$_2$(Cu$_{1-x}$Zn$_x$)$_3$O$_{6.9}$.
$x$ is the nominal zinc concentration and
$T_c$ is the superconducting
transition temperature. $A$ and $B$ are
extracted from a linear fit of the resistivity ($130<T<200$~K).
The scattering rate $\Gamma_{\rho}$ is estimated via the resistivity using
Eq.~(1) with
$\omega_p=1.3$~eV and
$\rho_0(x=0)=1~\mu \Omega$~cm.}
\begin{center}
\begin{tabular}{ccccc}
$x$ & $T_c$ & $A$ & $B$ & $\Gamma_{\rho}/T_{c0}$ \\
(\%) & (K) & ($\mu\Omega$~cm) & ($\mu\Omega$~cm~K$^{-1}$) & ($\hbar/k_B$)\\
\hline
pure  & 93.6 & -14.3 & 0.95 & $<$ 0.014 \\
0.6  & 89.2 & -6.0  & 1.00 & 0.13 \\
2  & 80.0 & 12.9  & 0.94 & 0.4 \\
3  & 74.6 & 22.9  & 1.07  & 0.54 \\
\end{tabular}
\end{center}
\end{table}

\begin{figure}[t]
\resizebox{\columnwidth}{!}{\includegraphics*{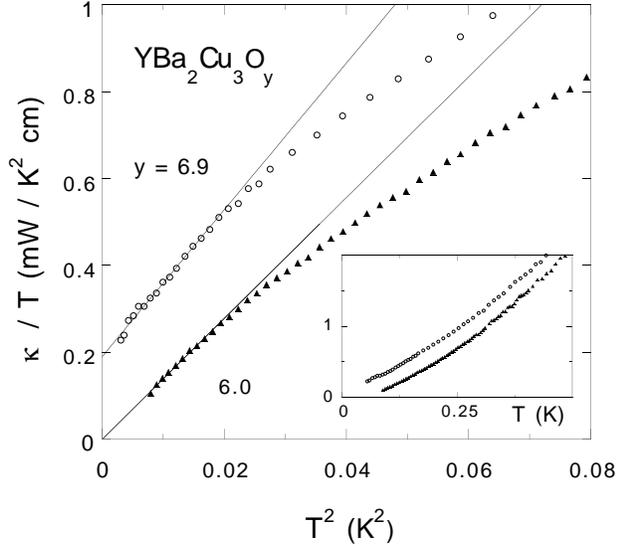}}
\caption{a-axis thermal conductivity of the two
YBa$_2$Cu$_3$O$_y$ crystals, one superconducting ($y=6.9$; circles) and one
insulating
($y=6.0$; triangles).
Main panel: $\kappa/T$ vs $T^2$; lines are fits to $a+bT^2$ for
$T <$~0.15~K.
Inset: $\kappa/T$ vs $T$.
}
\end{figure}

where
$\omega_p=\sqrt{4\pi n e^2 / m^*}$ is
the Drude plasma frequency and
$\rho_0(x=0)$ is the resistivity of the pure crystal at $T=0$.
The latter is estimated via the
microwave
conductivity, from which
the mean free path is known to increase by
$\simeq 100$ in going from 100~K to $\simeq$~10~K
in high-quality untwinned crystals
\cite{Bonn}.
Since $\rho_a$(100 K)~$\simeq 75~\mu \Omega$~cm, then
$\rho_0(x=0)<1~\mu\Omega$~cm.
With $\omega_p=1.3$~eV ($a$-axis) \cite{Timusk}, one gets
the scattering rates listed in Table I, in units of
$T_{c0}$=$T_c(x=0)$.
Note that {\it 3\% of Zn causes a 40-fold
increase in $\Gamma$}.

In order to use $\kappa(T)$ as a probe of quasiparticle
behavior, the phonon
contribution must
be extracted reliably.
This can only be done by going to temperatures sufficiently low that the phonon
conductivity
$\kappa_{ph}$ has reached its
well-defined asymptotic $T^3$ dependence, given by

\begin{equation}\label{kinetictheory}
\kappa _{ph}=\frac{1}{3} \beta \left\langle {v_{ph}} \right\rangle \Lambda_0
T^3
\end{equation}

where $\beta$ is the phonon specific heat coefficient,
$\left\langle {v_{ph}} \right\rangle$
is a suitable average of the acoustic sound velocities,
and $\Lambda_0$ is the temperature-independent maximum phonon mean free path.
In non-magnetic insulators, acoustic phonons are the only
carriers of heat at low temperature and Eq.~(2) is well verified,
with
$\left\langle {v_{ph}} \right\rangle =v_L(2s^2+1)/(2s^3+1)$ in single crystals,
where $s=v_L/v_T$ is the ratio of longitudinal to transverse velocities
\cite{Thacher}.
In high-quality crystals, $\Lambda_0 = 2 \bar{w} / \sqrt{\pi}$, where $\bar{w}$
is the (geometric) mean width of a rectangular sample
\cite{Thacher}.

The simplest way of investigating the phonon contribution in
YBa$_2$Cu$_3$O$_y$ is to remove all electronic carriers by
setting $y \simeq 6.0$.
(Note that
antiferromagnetic magnons
are not expected to contribute at $T<1~$K,
since the acoustic spin-wave gap in
YBa$_2$Cu$_3$O$_{6.15}$ is
$\simeq$~100~K \cite{Shamoto}.)
The thermal conductivity of such an insulating crystal is shown in Fig.~1
(triangles).
As
seen
from the linear fit,
 $\kappa/T=a+bT^2$
below about 0.15~K,
with $a = 0$ and $b = 14$~mW~K$^{-4}$~cm$^{-1}$.
The first question of interest is: what happens when electronic carriers are
introduced?
The answer is provided by the
thermal conductivity of a well-oxygenated crystal, also shown in Fig.~1
(circles):
{\it electronic carriers contribute a definite linear term to
$\kappa(T)$}.
Applying the same fit as before yields
$a=0.19$~mW~K$^{-2}$~cm$^{-1}$ and
$b = 17$~mW~K$^{-4}$~cm$^{-1}$.

It must be emphasized that such an analysis is sound only when applied to the
asymptotic regime for $\kappa_{ph}$. To confirm that this
is indeed the case for $T < 150 $~mK in the insulating crystal, note that
$a=0$ {\it and} the magnitude of the cubic term is right, i.e. it corresponds
to a maximum mean
free path $\Lambda_0$ dictated by the mean crystal width $\bar{w}$.
Indeed, from Eq.~(2) using
$\beta=0.3-0.4$~mJ/K$^4$~mole
\cite{Fisher,Moler}
and
$\left\langle {v_{ph}} \right\rangle=4000$~m/s
($v_L \simeq 6000$~m/s, $v_T \simeq 3700$~m/s
\cite{Dominec}),
$\Lambda_0 = 270-360~\mu$m~$= 2 \bar{w} / \sqrt{\pi}$ (see Table~II).
(Note that $\beta$ and $\left\langle {v_{ph}} \right\rangle$, given here for $y
\simeq 6.9$,
could be slightly different for $y \simeq
6.0$ \cite{Moler,Dominec}).
So the phonon mean free path in the
$y=6.0$ sample
unambiguously reaches its maximum, boundary-limited value
at $\simeq$~0.15~K.
It is then reasonable to expect a very similar {\it phonon}
behavior in the $y=6.9$ sample, given its nominally
identical crystalline quality and surface quality, and its
comparable dimensions.
This is nicely borne out by the $\kappa/T$ data in Fig.~1:
the only difference between the two curves ($y=6.9$ and 6.0) is a rigid offset.
In such a well-defined context, the appearence of a linear
term upon introducing electronic carriers
is conclusive evidence for the existence
of zero-energy quasiparticles in
YBa$_2$Cu$_3$O$_{6.9}$ and, as a result,
it confirms a key feature of the basic theory
of transport in unconventional superconductors
\cite{HVW,Varma,Lee,Maki,Graf,Norman,Hirschfeld}.
In this connection, earlier claims of a residual electronic linear term
in $\kappa$
of YBa$_2$Cu$_3$O$_{7-\delta}$
were inconclusive, being all based on the same analysis as used here
but applied to arbitrary temperature regimes
(for a review, see \cite{Uher}).

\begin{table}[hb]
\caption{
Parameters used in fitting
$\kappa/T$ to $a+bT^2$, where $a = \kappa_{0}/T$
is the electronic residual linear
term and
$b = \kappa_{ph}/T^3$
is the asymptotic phonon $T^3$ term.
$\bar{w}$ is the mean sample width and
$\Lambda_0$ is calculated from Eq.~(2) using
$\beta=0.3-0.4$~mJ/K$^4$~mole
and
$\left\langle {v_{ph}} \right\rangle=4000$~m/s.
}
\begin{center}
\begin{tabular}{ccccc}
sample & $\bar{w}$ & $\kappa_{0}/T$ & $\kappa_{ph}/T^3$ &
$\sqrt{\pi}\Lambda_0/2\bar{w}$ \\
 (\%)  & ($\mu$m) & (mW~K$^{-2}$~cm$^{-1}$) & (mW~K$^{-4}$~cm$^{-1}$) & \\
\hline
pure  & 252 & 0.19~$\pm$~0.03 & 17~$\pm$~2 & 1.2~-~1.6 \\
0.6\% & 242 & 0.17~$\pm$~0.04 & 11~$\pm$~2 & 0.8~-~1.0 \\
2\%   & 177 & 0.25~$\pm$~0.07 & 7~$\pm$~3  & 0.7~-~0.9\\
3\%   & 238 & 0.20~$\pm$~0.05 & 8~$\pm$~3  & 0.6~-~0.8 \\
\hline
deox  & 315 & 0.00~$\pm$~0.01 & 14~$\pm$~2 & 0.8~-~1.0 \\
\end{tabular}
\end{center}
\end{table}

\begin{figure}[t]
\resizebox{\columnwidth}{!}{\includegraphics*{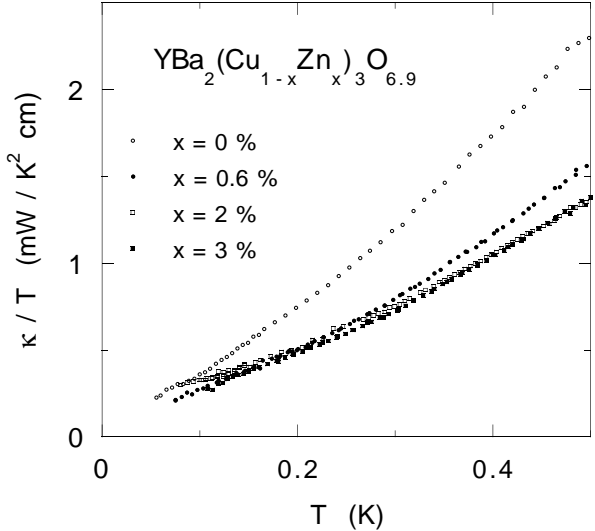}}
\caption{$a$-axis thermal conductivity
of the four Zn-doped crystals, plotted as $\kappa/T$ vs $T$.
}
\end{figure}

Having established the existence of a residual normal fluid in
YBa$_2$Cu$_3$O$_{6.9}$,
the next question is that of universality.
This is addressed by looking at
concentrations of Zn
such that $\Gamma$
ranges from $< 0.014$ up to 0.54~$T_c$.
The thermal conductivity of
YBa$_2$(Cu$_{1-x}$Zn$_x$)$_3$O$_{6.9}$
is shown in Fig.~2, where
it is apparent
that
$\kappa$ is unaffected by the variation in $\Gamma$
at $T\simeq0.1$~K, where
the heat is carried predominently
by quasiparticles
(cf. Fig.~1).
In other words, {\it transport by the residual normal fluid is universal}.

The $T\rightarrow0$ limit of
$\kappa/T$ is obtained from a fit to $a+bT^2$ limited to $T<150$~mK, as
applied earlier, which yields the values for $a=\kappa_{0}/T$ and $b$ listed in
Table~II.
Note that the ratio $\Lambda_0/(2 \bar{w} / \sqrt{\pi}) \simeq  1$ for all
crystals,
proving that the asymptotic phonon regime
was reached in all cases.
(The somewhat larger ratio for the pure sample is intriguing -- further work is
needed to
elucidate this.)
As seen from
a plot of
$\kappa_0/T$ versus $\Gamma$,
shown in Fig.~3,
these values are consistent with a universal linear term of
$0.19$~mW~K$^{-2}$~cm$^{-1}$.
Note, however, that
the error bars on the values of $a$ and $b$ in Table~II are fairly large,
because they
combine uncertainties on the geometric factors (largest for the rather short
2\% sample)
and on the fit, which is limited to a small temperature range (smallest for the
3\% sample).
One way of eliminating the uncertainty on the geometric factor
is to use the resistivity data obtained with the same contacts.
Indeed, by fixing $B=1.03~\mu \Omega$~cm~K$^{-1}$
for all samples, thereby imposing the reasonable constraint that
the inelastic scattering is not affected by small levels of Zn, one can correct
$\kappa_0/T$ by multiplying it by $B(x)/1.03$. This yields the following
corrected values:
$0.17\pm 0.01$,
$0.17\pm 0.02$,
$0.23\pm 0.02$ and
$0.21\pm 0.04$~mW~K$^{-2}$~cm$^{-1}$,
for $x=0$, 0.6, 2 and 3\%, respectively.
These are plotted in the inset of Fig.~3 versus the similarly corrected
$\Gamma_{\rho}$.
The corrected plot with its smaller error bars no longer allows for a constant
linear term:
there is a small but definite upward slope, with a minimum growth of 30\% over
the range of
$\Gamma_{\rho}$
and a maximum growth of 55\%.
{}From this we conclude that while the residual linear term is universal, in
the sense that a
10-fold increase in $\Gamma$ (from 0.014 to 0.13~$T_c$ in going from $x=0$ to
0.6\%) leaves
$\kappa_0/T$ unchanged, at
larger $\Gamma$ there is a slight increase, reaching approximately 40\% at
$\Gamma/T_c \simeq 0.6$.

\begin{figure}[t]
\resizebox{\columnwidth}{!}{\includegraphics*{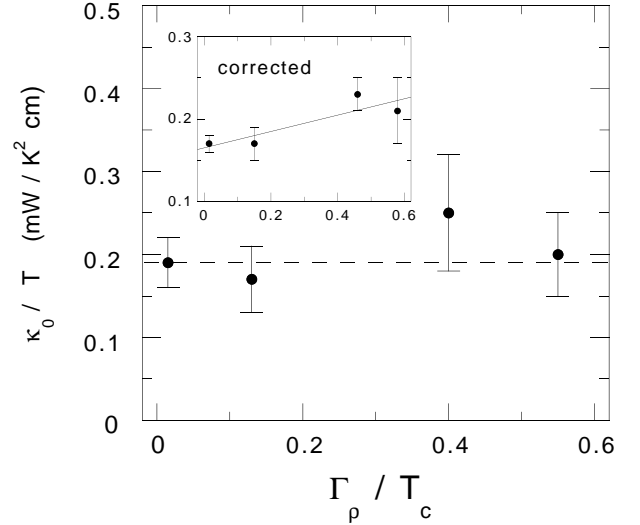}}
\caption{
Residual linear term vs scattering rate for the four crystals of
YBa$_2$(Cu$_{1-x}$Zn$_x$)$_3$O$_{6.9}$; the dashed line indicates a constant at
0.19
mW~K$^{-2}$~cm$^{-1}$.
Inset: same, but with corrected values (see text); the solid line is a
least-squares
fit.}
\end{figure}

Let us now compare our results with the theory of heat transport in
unconventional
superconductors \cite{Maki,Graf,Norman,Hirschfeld}.
The first point to emphasize is that universality is only expected for special
gap functions
with appropriate topology and symmetry. This is the case for
a pairing state of $d_{x^2-y^2}$ symmetry,
with line nodes at azimuthal angles $\phi=m \pi /4$ ($m=1,3,5,7$).
The $T=0$ limit of $\kappa/T$ along the $x$ (or $a$) direction is
\cite{Maki,Graf}

\begin{equation}
\frac{\kappa_{00}}{T} = L_0 \sigma_{00}
\rightarrow \frac{L_0 n e^2}{m^*} \frac{2\hbar}{\pi S}
= \frac{\hbar k_B^2 \omega_p^2}{6 e^2} \frac{1}{S}
\end{equation}

where $\sigma_{00}$ is the universal limit of charge conductivity
\cite{Lee,Graf},
$L_0=(\pi^2/3)(k_B/e)^2$
is the Sommerfeld value of the Lorenz number,
and
$S = |d\Delta(\phi)/d\phi|_{node}$
is the slope of the gap at the node
\cite{Graf}.
The topology of the excitation gap right at the node
(e.g. $\Delta(\phi) \sim \phi - \pi/4$)
determines
universality
and then the slope $S$
sets the magnitude of $\kappa_{00}/T$.
Note that a gap function of the right topology but of $s$-wave symmetry will
{\it not}
in general show universal behavior, so that
our observation of a universal $\kappa_0/T$ is strong support for a gap of
$d$-wave symmetry.

A quantitative comparison with the theory reinforces this conclusion. In the
simplified case
of the
standard $d$-wave gap $\Delta_0$cos$(2\phi)$, $S=2\Delta_0$.
Using available estimates of
$\hbar \omega_p$ and the gap maximum
$\Delta_0$, respectively equal to
1.3~eV \cite{Timusk} and
20~meV \cite{Geneva},
one gets

\begin{equation}
\frac{\kappa_{00}}{T} = 0.09~ \rm{mW K^{-2} cm^{-1}}
\end{equation}

which is remarkably close to the measured value of 0.19~mW~K$^{-2}$~cm$^{-1}$.
Given that the real gap will have more structure than a simple cos$(2 \phi)$
dependence, the factor 2 discrepancy
suggests that it actually rises from the node half as fast as in the simple
model.
This
in no way detracts from the conclusion that a
(generalized) {\it $d$-wave gap is in excellent quantitative agreement with the
universal heat
conduction observed in YBa$_2$Cu$_3$O$_{6.9}$}.

The second point to consider in a comparison with the theory is the fact that
universality is only achieved when $\hbar\gamma<<\Delta_0$, where $\hbar\gamma$
is the
bandwidth of impurity bound states responsible for the zero-energy
excitations
\cite{Graf}.
The bandwidth grows with $\Gamma$ in a way which depends very
strongly on the
scattering phase shift $\delta_0$.
It is largest in the limit of unitarity scattering,
$\delta_0 = \pi/2$,
where $\hbar \gamma \sim
\sqrt{\pi \Delta_0 \hbar \Gamma/2}$ \cite{Graf}.
For the pure and 3\% samples,
with $\Gamma_{\rho}/T_{c0} = 0.014$ and 0.54,
this gives $\hbar\gamma / \Delta_0 \simeq$
0.1 and
0.6,
respectively (for $\Delta_0 \simeq 20$~meV~=~2.5~$k_B T_{c0}$).
Thus we expect
deviations from universality
for the samples with high Zn doping.
Quantitatively,
the dependence of
$\kappa_{00}/T$ on
$\Gamma$ was calculated by Sun and Maki \cite{Maki}, who
find a monotonic increase, which gets to be a factor 1.9 at
$\Gamma/T_{c0}=0.54$ (see also
Ref.~17).
Such a large increase is incompatible
with the data
(see Fig.~3).
On the other hand, a 40\% growth in the residual linear term,
consistent with the data, would agree with the
calculation if $\Gamma \simeq \Gamma_{\rho}/2$,
namely 0.3~$T_{c0}$
for the 3\% sample
instead of 0.54~$T_{c0}$.
Interestingly, this is the $\Gamma$ one deduces self-consistently from the
theory \cite{Sun},
based on the
measured $T_c$ suppression.
Note, however, that accounting for a smaller $\Gamma$ in terms of a smaller
``effective''
$\omega_p^2$
in Eq.~(1) leads to an even smaller
$\kappa_{00}/T$ from Eq.~(3).

These minor discrepancies notwithstanding, one of the main implications of the
good agreement
with the theory is that impurity scattering in the cuprates is well-described
by a phase
shift very close to $\pi/2$ \cite{Graf},
something which has been assumed often but rarely verified.
Nonetheless, a proper interpretation of the data should include the possibility
of a small departure from the unitarity limit \cite{Hirschfeld}.
This would lower $\gamma$, making it easier to satisfy the condition
$\hbar \gamma << \Delta_0$, and possibly to account for the weak
variation in
$\kappa_0/T$.

The present results have implications for other properties of
YBa$_2$Cu$_3$O$_{6.9}$, such as specific heat \cite{Fisher,Moler}
and microwave conductivity \cite{Zhang}. It has not yet been
possible to probe the residual normal fluid reliably via these properties, but
the upper
bounds imposed by the data so far are consistent with the behavior
predicted on the basis of the thermal conductivity data reported here.

In summary, we presented the first observation of universality in the transport
properties
of a
superconductor.
A residual linear term $\kappa_{0}/T = 0.19$~mW~K$^{-2}$~cm$^{-1}$ is clearly
resolved
in the thermal conductivity of YBa$_2$Cu$_3$O$_{6.9}$
and attributed to electronic carriers.
The observation of this residual normal fluid is a powerful validation of the
basic theory
of impurity scattering in unconventional superconductors.
The fact that $\kappa_0/T$ is universal, i.e.
virtually
unaffected by changes in the impurity scattering rate, strongly confirms the
gap has having
$d$-wave symmetry.
However, from the magnitude of $\kappa_0/T$, it appears that the gap rises more
slowly at the
nodes than described by the standard function $\Delta_0$cos$(2\phi)$ with
$\Delta_0 = 2.5~k_B
T_{c0}$.

We thank John Berlinsky, Brett Ellman, Matthias Graf, Peter Hirschfeld and
Catherine Kallin
for stimulating
discussions.
This work was funded by NSERC of Canada, FCAR of Qu\'ebec and
the Canadian Institute for Advanced Research. L.T.
acknowledges the support of the Alfred P. Sloan Foundation.

\end{document}